\documentclass[12pt,draft,onecolumn]{IEEEtran}

\usepackage{srcltx}
\usepackage{cite}
\usepackage[final]{graphicx}
\usepackage[]{graphics}
\usepackage{caption}
\usepackage{subcaption}
\usepackage{amssymb}
\usepackage{epstopdf}
\usepackage{array}
\usepackage{epsfig}
\usepackage{balance}
\usepackage{float}

\usepackage[cmex10]{amsmath}

\begin{document}
\title{Low Complexity Two-Stage Soft/Hard  Decoders}

\author{Guido Montorsi and Farbod Kayhan
\thanks{Parts of this paper has been presented in the 3$^{rd}$ IEEE international Black Sea Conference on communications and networking under a paper titled "Analog Digital Belief Propagation in two Stage Decoding Systems with a Hard Decoding Stage", Constanta, May 2015. The paper won the best paper award of the conference.

Guido Montorsi are with the Department of Electronics and Telecommunication, Politecnico di Torino,
10129 Torino, Italy (email: guido.montorsi@polito.it). Farbod Kayhan is with Interdisciplinary
Centre  for  Security, Reliability  and  Trust (SnT), University of Luxembourg. (email:
farbod.kayhan@uni.lu). }}

\maketitle

\begin{abstract}
Next generation wireless systems will need higher spectral efficiency as the expected traffic volumes per unit bandwidth and dimension will inevitably grow. As a consequence, it is necessary to design coding schemes with performances close to the theoretical limits, having high flexibility and low complexity requirements at transmitter and receiver. In this paper, we point out some of the limitations of the Bit Interleaved Code Modulation (BICM) technique which is the state of the art adopted in several standards and then propose some new lower complexity alternatives. These low complexity alternatives are obtained by applying the recently introduced Analog Digital Belief Propagation (ADBP) algorithm to a two stage encoding scheme embedding a hard decoding stage. First we show that for PAM$^2$ type constellations over the AWGN channel, the performance loss caused by using a hard decoded stage for all modulation bits  except the two least protected is negligible. Next, we consider the application of two stage decoders to more challenging Rician channels, showing that in this case the number of bits needed to be soft decoded depends on the Rician factor and increases to a maximum of three bits per dimension for the Rayleigh channel. Finally, we apply the ADBP algorithm to further reduce the detection and decoding complexity.
\end{abstract}

\IEEEpeerreviewmaketitle

\section{Introduction}
In view of the growing demand for spectral efficiency in wireless systems, coding and modulation
design has received considerable attention in the past few decades. In particular, BICM employing
soft iterative decoders has been adopted in several standards, as it provides a performance very
close to the theoretical limits \cite{BICM}. Soft iterative decoding schemes play a central role in
this achievement. However, the complexity at the receiver increases considerably by increasing the
modulation cardinality and code length. In this paper we consider an alternative solution based on
multistage coding which significantly reduces the receiver complexity without compromising the
performance.

Multilevel coding (MLC) as a method to jointly optimize coding and modulation has been introduced
by Imai and Hirakawa \cite{Imai_Hirakawa}. At the receiver, each coding level is decoded
individually considering the decision of the prior stage. Such decoding scheme is usually referred
to as multistage decoder (MSD) in the literature. It is well known that by employing a hard
decoding at all stages a large loss up to 2 dB is expected (see for example
\cite{Wachsmann_MLC_Hard}). On the other hand, applying a maximum likelihood decoder or soft
decoding at all stages increases the complexity of the decoder making it less practical in
comparison with the BICM technique \cite{BICM}.

Despite the vast amount of research on the multilevel coding, few works are devoted to the decoding
techniques. The main results of multistage schemes employing hard decoding stages are presented by
Wachsmann et al. in two successive papers \cite{Wachsmann_MLC_Hard} and \cite{Wachsmann_MLC}. In
these papers, the authors consider a multistage coding technique over the AWGN channel and analyze
the complexity of the decoder. They show that the decoder complexity can be reduced substantially
by using the hard decisions in several decoding stages. In particular, they show that if only the
lowest level coded bit is soft-decoded the performance degradation can be kept quite small.
However, as we will see, this is true only if the code rates are chosen carefully and therefore the
coding scheme is not very flexible.

The authors in \cite{Philippa} use a hard decoder in all stages. In order to compensate for the
large loss in performance, they consider the possibility to pass the reliable information to both
previous and subsequent decoders. This strategy becomes very complex in the realistic design and is
out of interest in our research, as we would like to have sequential MSD decoder. Similarly, in
\cite{IHID}, the authors accept the loss due to the hard decoder and try to optimize the decoder by
proposing an improved hard iterative decoding (IHID) algorithm which efficiently terminates hard
iterations. In \cite{DRM}, the authors analyze different implementations of the MSD utilizing
hard-decision with several metrics. The main idea is to change the channel model from one stage to
the other. In particular, they avoid passing the unreliable information from one decoding stage to
the subsequent decoder in order to avoid the propagation of the errors. Therefore, the channel for
the subsequent decoder is a mixture of Gaussian and erasure channel.

In this paper we propose a \textbf{two level} coding scheme in which only few bits, associated to
the first level encoder, are soft decoded. We show that over AWGN channel choosing two soft decoded
bits, instead of only one bit as was proposed in \cite{Wachsmann_MLC_Hard}, allows for a  more
flexible coding design keeping the performance degradation negligible. We further show that over
the Rician channel, the number of bits that need to be softly decoded depends on the Rician factor
and increases to 3 bits per dimension for the Rayleigh channel.

Then, we compare the complexity and performance of this coding technique with the state of the art
BICM scheme. We show that by considering only two stages, one can reduce the design complexity and
latency but still fully benefit of the complexity reduction at the receiver and flexibility at the
transmitter. Moreover, following the results in \cite{Forney}, we show that, when the employed
modulations are subset of lattices like PAM modulations, the channel for the first layer decoder
can be modeled as a wrapped additive white Gaussian noise (WAWGN) channel. This property can be
used to further simplify the computation of the log-likelihood ratio (LLR) in the soft decoding
stage and make the normalized complexity of the proposed decoding scheme actually decreasing with
the spectral efficiency.

Finally, observing that the model of the wrapped Gaussian channel for the soft encoded bits is
perfectly matched to the model used for the derivation of ADBP \cite{ADBP_lett}, we propose an even
simpler two stage decoding scheme using ADBP for the soft decoded stage and hard decoding for the
second stage.

We organize the paper as follows. In section \ref{sec:MLC} we shortly review the MLC scheme and the
corresponding multistage decoder. In section \ref{sec:complexity} we compute the decoding
complexity of a two level coding and compare it with that of BICM approach. In section
\ref{sec:Results1} we compare the performance and flexibility of MLC scheme with the BICM by
calculating the loss from the Shannon capacity for each system. Performance and flexibility of the proposed two level scheme over Rician and Rayleigh channels are studied in section \ref{sec:Rice}. In section \ref{sec:LLRsimp} we
discuss the simplification of the LLR computations for the soft decoder in the first stage. We
shortly introduce the ADBP algorithm in section \ref{sec:ADBP} and show how it can be embedded in
our two level MLC scheme. The  simulation results for our system and comparison with the BICM
performance is provided in section \ref{sec:simul}. Finally, we conclude the paper in section
\ref{sec:Conclusion} and summarize the main results.

\section{Multi level encoding and multistage decoding}\label{sec:MLC}
In single stage encoder, the information bits are encoded by a binary encoder and mapped, possibly
after an interleaver, to a constellation set with cardinality $M=2^m$. In MLC schemes with $L$
levels, the information block is split into $L$ subsets that are independently encoded by different
binary encoders \cite{Imai_Hirakawa}. The encoders may have different rates and the output of the
$i^{th}$ encoder is associated to a subset of length $m_i$ of the binary label of the
constellation, so that $m=\sum_{i=1}^{L}m_i$. We assume that the size of the $i^{th}$ level
codeword is $m_iN$ and we define $M_i= 2^{m_i}$. At the receiver, each level is decoded
individually, starting from the lowest level, and taking into account decisions of prior decoding
stages. It is known that this scheme can indeed achieve the channel capacity
\cite{Imai_Hirakawa,Wachsmann_MLC,Forney}. More precisely, let $B=(B_1, B_2,...,B_L)$ denotes the
constellation binary label of length $m$, where each $B_i$ is the bit sequence of length $m_i$
related to the $i^{th}$ encoder. We denote any realization of the random variable $B$ by lower case
$b=(b_1,b_2,...,b_L)$. The transmitted signal $X$ is associated through a mapping $X(B)$ to each
label $B$. The received signal is then  $Y=X(B)+Z$ where $Z$ is the AWGN. Then the capacity of the
MLC coding system can be written as
\begin{equation}
C=\sum_{i=1}^{L} I(Y;B_i|\bar{B}_i) = \sum_{i=1}^{L} H(Y|\bar{B}_{i})-H(Y|\bar{B}_{i+1}),
\end{equation}
where $I(.;.)$ is the mutual information function and $\bar{B}_i \triangleq (B_1,...,B_{i-1})$.
Notice that in a practical implementation each constituent code must be designed to match the
corresponding stage capacity and therefore the design is complicate and usually not flexible.

\subsection{Two stage MLC scheme (2SD-SH)}
From the encoding perspective there is no apparent advantage using MLC,  indeed, the presence of
multiple codes requires the careful design of each encoding stage  increasing the design
complexity. Furthermore, the need of splitting the information block size into smaller blocks leads
to a reduction of the block size of each individual encoder and hence increasing the latency. As a
consequence, MLC are seldom used in the applications where there is no need to divide the
information bits into sub blocks.

On the other hand, as it was shown in \cite{Wachsmann_MLC_Hard}, at the receiver it is possible to
reduce the decoder complexity by using  hard decoding in almost all stages without loss in
performance. We will show that exploiting this possibility we can construct decoding schemes with
performance and flexibility comparable to that of BICM scheme, but with lower complexity.

\begin{figure}[tb]
	\begin{center}
	\includegraphics[angle=-90,width=\the\hsize]{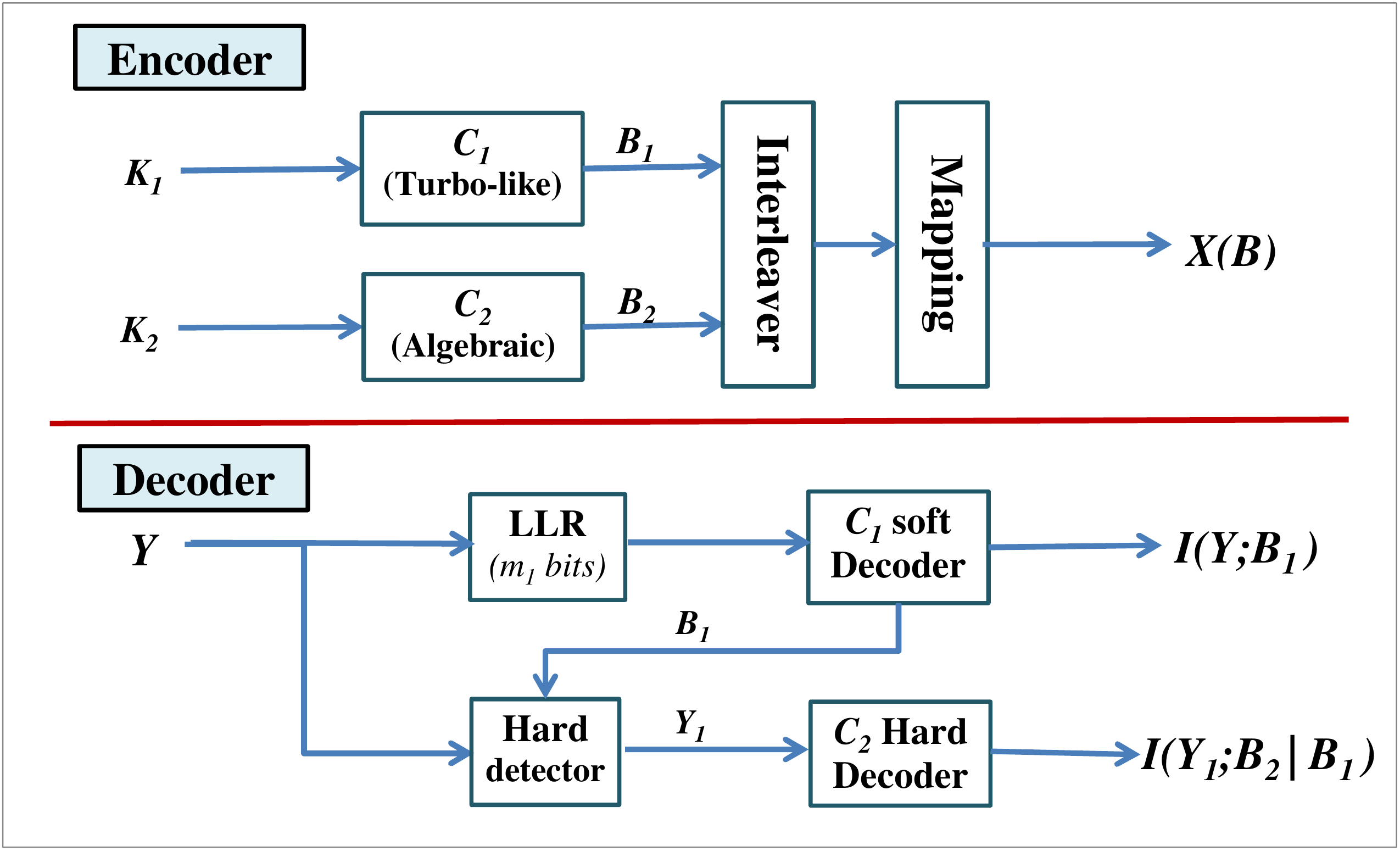}
	\caption {\label{Fig:MLC_enc_dec} Block diagram of the considered MLC encoder and the corresponding MSD scheme (2SD-SH).}
	\end{center}
\end{figure}

In Fig. \ref{Fig:MLC_enc_dec} we report the general block diagram of the considered two stage MLC
and the corresponding MSD receiver embedding a hard decoding stage, named 2SD-SH in the following.
The information block is divided in two blocks and encoded separately by two stages. The first
stage  encodes $m_1$ bits by using a powerful capacity achieving turbo-like encoder, while the
second one encodes the remaining $m_2=m-m_1$ bits with a simpler algebraic encoder. At the
receiver, a soft-input (using bit-LLR, like BICM) iterative receiver is used to decode the first
stage (B). The output of the first stage is provided to the next stage (H), which takes hard
decisions on the observations and then uses a simpler algebraic decoder (hard decoder).

\subsection{Constellation Mapping}\label{subsec:labeling}
For the 2SD-SH schemes it is important to optimize the constellation set labeling. The goal is to
minimize the bit error probability of the sub-constellation sets assuming that the soft bits have
been successfully decoded from previous stage. As it was shown in \cite{Wachsmann_MLC}, the optimal
mapping is a mixture of set partitioning (for the two stages) and Gray mapping within each stage.
The Gray mapping within each stage minimizes the loss of the BICM soft decoder on the first stage
and the bit error probability (BER) for the hard decoded stage. For more details we refer the
readers to \cite{Wachsmann_MLC} and the references therein. In this paper we focus only on
$2^{m}$-QAM constellations with even $m$. In this case, the mapping design is rather
straightforward and the optimization problem can be reduced to the one dimensional $2^{m/2}$-PAM
constellation.

\section{MLC vs BICM: Complexity}\label{sec:complexity}
As we have mentioned, BICM is the current solution adopted in several standards for flexible high
spectral efficiency transmission \cite{BICM}. The receiver complexity comparison of the MLC scheme
of Fig. \ref{Fig:MLC_enc_dec} with respect to BICM scheme has been presented in \cite{blacksea}.
For the sake of completeness, we briefly report the results here again. Notice that in BICM
receivers there are no iterations between the detector and the decoder and usually the decoding
algorithm is stopped after reaching a fixed maximum number of iterations $I_l$. For simplicity, we
have not considered the effect of $I_l$ in the following computations. Let
$\mathcal{C}_{\text{soft}}$ and $\mathcal{C}_{\text{hard}}$ denote respectively the complexity per
decoded information bit of the soft and hard decoder. We also denote by $\mathcal{C}_{\text{LLR}}$
and by $\mathcal{C}_{\text{HD}}$ the complexity of the bit LLR computer and hard decision block,
normalized to the number of constellation points $M$.

The total complexity of a  2SD-SH decoder, normalized to the information block length $K$, can be
evaluated as
\begin{eqnarray}\label{eq:complexity}
\mathcal{C}_{\text{MSD}} &=&  \mathcal{C}_{\text{soft}} \frac{K_1}{K} +
\mathcal{C}_{\text{hard}} \frac{K_2}{K} + \mathcal{C}_{\text{LLR}}\frac{N}{K}2^m + \mathcal{C}_{\text{HD}}\frac{N}{K}2^m \nonumber  \\
&=& x\mathcal{C}_{\text{soft}} +  (1-x)\mathcal{C}_{\text{hard}} + \frac{2^m}{mR_c} (\mathcal{C}_{\text{LLR}}+ \mathcal{C}_{\text{HD}})\label{eq:comp}
\end{eqnarray}
where $x \triangleq \frac{m_1 R_1}{mR_c}$ is the ratio between the throughput associated to the
soft stage and the total decoder throughput. The complexity of a BICM decoder can be obtained from
(\ref{eq:complexity}) by setting $x=1$ and removing the complexity of the hard detector, i.e.:
\begin{equation}\label{eq:complexity_BICM}
\mathcal{C}_{\text{BICM}} = \mathcal{C}_{\text{soft}} + \frac{2^m}{mR_c} \mathcal{C}_{\text{LLR}}.
\end{equation}

In particular, setting a fixed value for $m_1$ the complexity $\mathcal{C}_{\text{MSD}}$ for large
$m$ will be dominated by the complexity of the \emph{hard decoder} and by that of the LLR computer,
that is still exponential with $m$.

Notice that the computational complexity of $ \mathcal{C}_{\text{LLR}}$ exponentially increases as
a function of $m$ and hence also that of spectral efficiency. This term is present in both
(\ref{eq:complexity_BICM}) and (\ref{eq:complexity}). As we will see in section \ref{sec:ADBP}, for
the MLC we can use a ADBP algorithm which simplifies the LLR computations and hence further reduces
the receiver complexity for the 2SD-SH scheme.

\section{MLC and BICM: Performance and Flexibility}\label{sec:Results1}
In order to perform a comparison between 2SD-SH and BICM in terms of performance and flexibility we
evaluate the mutual information (MI) for the two layer MLC presented in Fig. \ref{Fig:MLC_enc_dec}.
The MI computation can be split into two parts: $I(Y;B_1)$ and $I(Y_1;B_2 | B_1)$, where $Y_1$ is
the random variable at the output of the hard detector for the second stage. The hard detector
converts the second level channel into an equivalent Binary Symmetric Channel (BSC) and we have:
\begin{equation}
I(Y_1;B_2 | B_1) = m_2N (1-H(p)),
\end{equation}
where $p$ is the BER of the equivalent BSC channel as seen from the algebraic decoder. For more
details on the MI computation for MLC with hard decision stages we refer the reader to
\cite{Wachsmann_MLC_Hard} and \cite{Wachsmann_MLC}. Notice that BICM can be considered as a special
case of the two stage scheme by choosing $m_1=m$ and $m_2=0$.

\begin{figure}
    \centering
    \vspace*{-0.5in}
    \begin{subfigure}[b]{0.45\textwidth}
        \includegraphics[width=\textwidth]{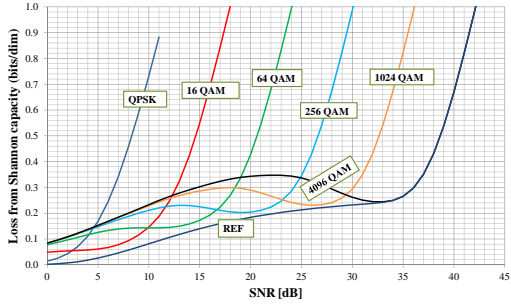}
        \vspace*{-0.6in}
        \caption{Loss from the capacity for BICM}
        \label{fig:gull}
    \end{subfigure}
    \hfill
    \vspace*{-0.2in}
    \begin{subfigure}[b]{0.45\textwidth}
        \includegraphics[width=\textwidth]{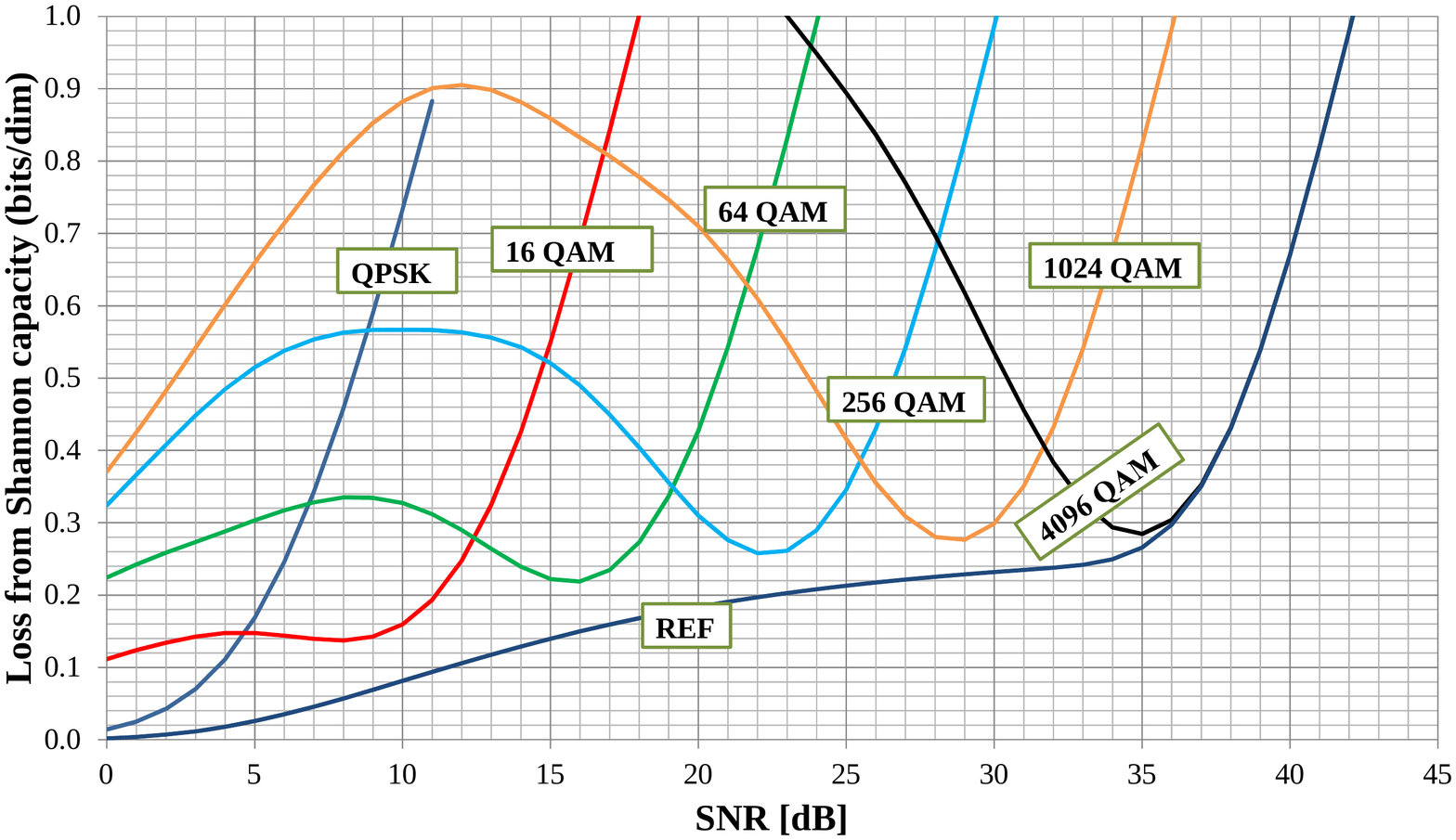}\\
          \vspace*{-0.5in}
        \caption{2SD-SH with one soft bit decoding per dimension }
        \label{fig:tiger}
    \end{subfigure}
    \hfill
    \begin{subfigure}[b]{0.45\textwidth}
        \includegraphics[width=\textwidth]{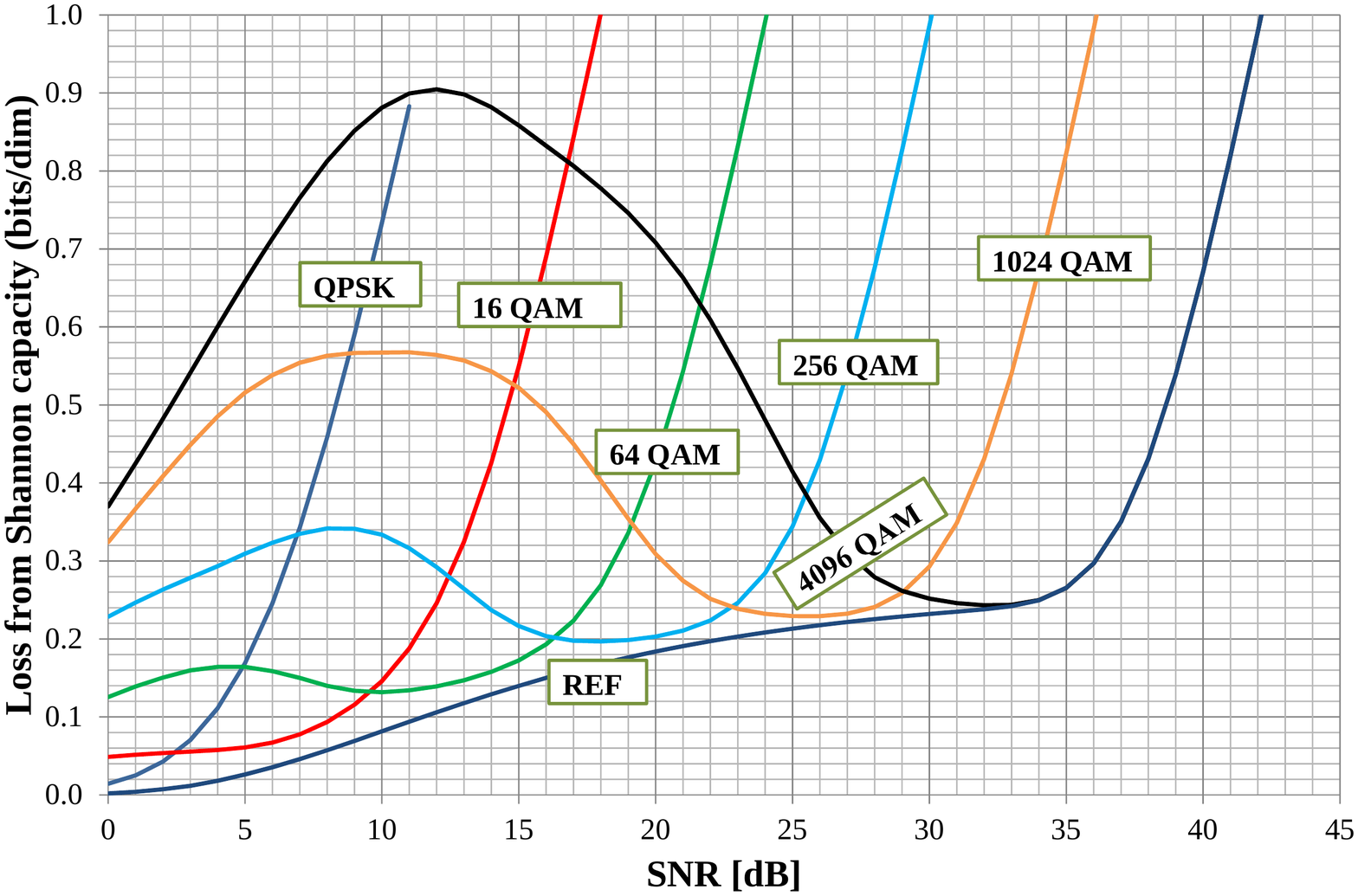}
          \vspace*{-0.5in}
        \caption{2SD-SH with two soft bits decoding per dimension }
        \label{fig:mouse}
    \end{subfigure}  \hfill
        \begin{subfigure}[b]{0.45\textwidth}
        \includegraphics[width=\textwidth]{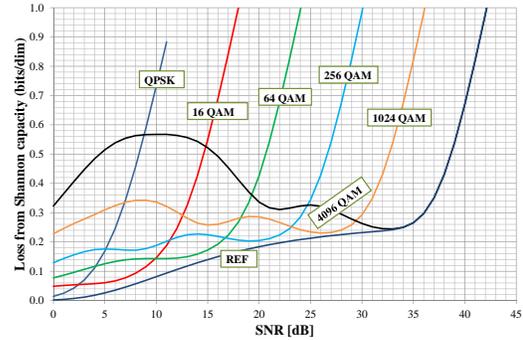}
          \vspace*{-0.5in}
        \caption{2SD-SH with three soft bits decoding per dimension }
        \label{fig:mouse}
    \end{subfigure}  \hfill
    \caption{Loss from the capacity over the AWGN channel for BICM and 2SD-SH schemes with various number of soft decoded bits per dimension.}\label{fig:AWGN}
\end{figure}

In Fig. \ref{fig:AWGN}(a), for reference, we present the MI results for the BICM system for several
$M$-QAM constellations. Notice that in this figure we report the \emph{loss}, in bits per
dimension, with respect to the Shannon limit $\log(1+SNR)$ instead of the actual MI. BICM, with
Gray mapping, has performance very close to the (QAM) limits, provided that constellation sets are
used in a correct range of SNR. The reference curve (REF) is the mutual information loss from the
capacity of the 16384-QAM constellation. This loss is due to the shaping loss of QAM, that
asymptotically amounts to 0.25 bits per dimension. Same reference curve is also used in Figs.
\ref{fig:RICE} and \ref{fig:RAY}. The optimal range of SNR for a constellation of size $m$
corresponds to spectral efficiencies $R_cm$ in the range $m-1.5,m-0.5$ and thus requires code rates
in the range $R_c\in[\frac{m-1.5}{m},\frac{m-0.5}{m}]$.
In Fig. \ref{fig:AWGN}(b) we present the MI results for a two stage system with \textbf{one} soft
decoded  bit ($m_1=1$) per dimension for several $M$-QAM constellations. As it was  mentioned in
\cite{Wachsmann_MLC_Hard}, the loss from the Shannon capacity is small if the code rate is chosen
carefully. However, it can be observed that, for each given constellation cardinality $M$, there is
a rather small range of SNRs where this loss is small. This will reduce considerably the
flexibility of the code design and does not allow to construct good encoding schemes for all
spectral efficiencies. This problem can be solved by increasing the number of soft decoded bits to
$m_1=2$. In Fig. \ref{fig:AWGN}(c) we show the capacity curves for such a system. As it can be
seen, not only the loss from the capacity decreases, but the capacity curves for each given $M$ are
much more flat at their minimum. Therefore, there is a larger range of ``optimal'' SNR for each
constellation set, so that for each SNR there is an MLC solution yielding negligible loss. Indeed,
the optimal range of SNR for a constellation of size $2^m$ corresponds to the same optimal range of
spectral efficiencies found for BICM ($[m-1.5,m-0.5]$). Given that in this case we use  $m_1 = 2$,
and that the algebraic coder rate $R_2$ is approximately one, the required range of coding rates
for the soft encoded stage is $R_1 \in [\frac{0.5}{2},\frac{1.5}{2}]$, \emph{independently} from
the constellation size $M$. Finally, in Fig. \ref{fig:AWGN}(d) we plot the similar results for the
case $m_1=3$. The results show that no additional gain (both in terms of performance and
flexibility) can be achieved by increasing the number of soft decoded bits and therefore $m_1=2$ is
the optimal trade off between the complexity, performance and flexibility of the MLC scheme over
the AWGN channel. As we will see later, this may not be true for different channel models.   	

To conclude, a two stage MLC with two soft decoded bits and a hard decoding stage for the remaining
$m_2=m-2$  is less complex than BICM (see equation (\ref{eq:comp})). Moreover, this scheme is more
flexible and scalable as all desired spectral efficiencies can be obtained coupling a suitable
modulation with a flexible turbo-like encoder with rate in the fixed range $R_1=[1/4,3/4]$ instead
of $[\frac{m-1.5}{m},\frac{m-0.5}{m}]$ as required by BICM approach.

\subsection{Performance of MLC with two stages}
	\begin{figure}[tbh]
	    \begin{center}
	    \vspace*{-0.6in}
	        \includegraphics[angle=0,width=\the\hsize]{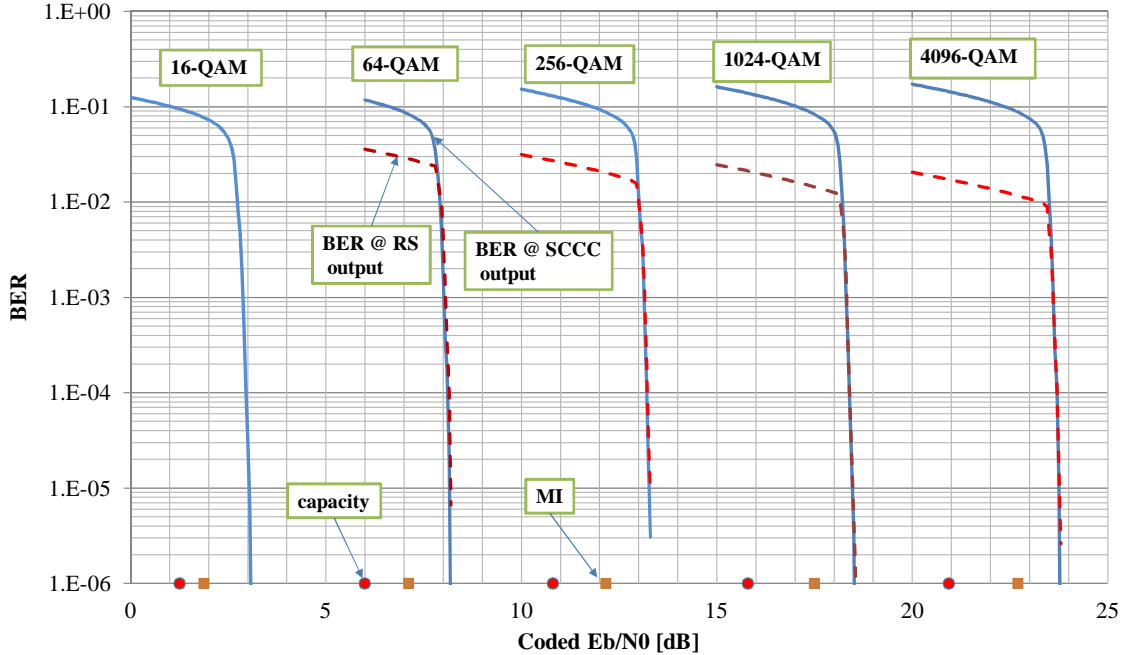}
	        \vspace*{-0.5in}
	    \caption {\label{Fig:BER_2bitsoft}The BER simulations for proposed two level code with 2 soft decoded bits per dimension for several QAM constellation cardinalities.}
	       \end{center}
	\end{figure}
In order to validate the conclusion of the previous section using  practical encoding scheme, we
have simulated the bit error rate (BER) over the  AWGN channel of a two level MLC scheme
constructed using an SCCC encoder for the first stage and a Reed-Solomon code for the second one.
In Fig. \ref{Fig:BER_2bitsoft} we present the BER simulations for several values of $M$. A SCCC
code with rate $R_1=2/3$ is used in all cases. The Reed-Solomon parameters are $q=8$ and $t=2$
($R_2=251/255$). The total code rates for each given $m$ can be calculated as
$$R_c = \frac{1}{m}(m_1R_1+m_2R_2).$$

For each given $M$ we report the BER for both soft decoder (blue solid lines) and hard decoders (red dashed lines). Notice that since we are using 2 soft bits per dimension in the first stage, for 16-QAM constellation all bits are soft decoded and the second stage does not exist in this case. Also it worth to notice that the performance is not degraded after the second stage, which is consistent with the mutual information results in the previous section. In the same figure we also indicate the Shannon capacity (red circles on horizontal axis). The square points at
$10^{-6}$ on the left of each curve corresponds to the maximum MI that can be achieved using the QAM constellations. This indicates that our coding scheme performs within 1 dB from the predicted limits.

\section{Performance of two stage decoder with QAM constellations over Rician channel} \label{sec:Rice}
In this section we will analyze the effect of a Rice fading on the performance of the two stage
decoders with soft/hard stages (2SD-SH). The main drawback of the 2SD-SH scheme is that the second
stage is weakly protected. A possible problem which may arise is then the robustness of this
encoding system to fading environments and more realistic scenarios. In this section we consider a
Rice channel with factor $K$ and investigate the maximum number of bits that can be hard decoded
for a negligible capacity loss. The details for the MI computations are as in the last section.

\begin{figure}
    \centering
    \vspace*{-0.5in}
    \begin{subfigure}[b]{0.45\textwidth}
        \includegraphics[width=\textwidth]{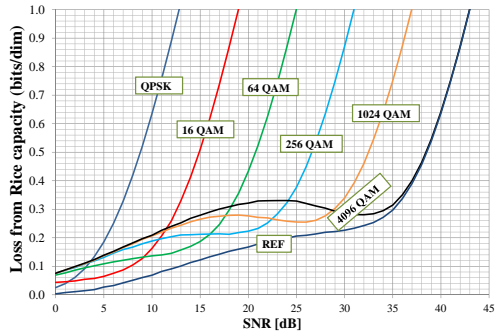}
        \vspace*{-0.5in}
        \caption{Loss from the capacity for BICM}
        \label{fig:gull}
    \end{subfigure}
    \hfill
    \vspace*{-0.1in}
    \begin{subfigure}[b]{0.45\textwidth}
        \includegraphics[width=\textwidth]{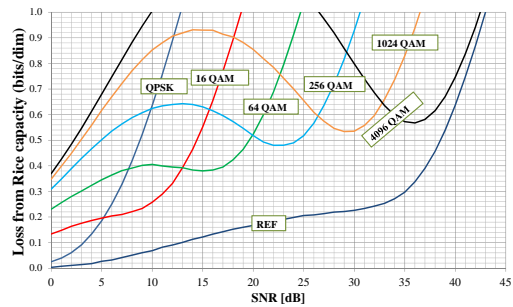}\\
          \vspace*{-0.55in}
        \caption{2SD-SH with one soft bit decoding per dimension }
        \label{fig:tiger}
    \end{subfigure}  \hfill
    \begin{subfigure}[b]{0.45\textwidth}
        \includegraphics[width=\textwidth]{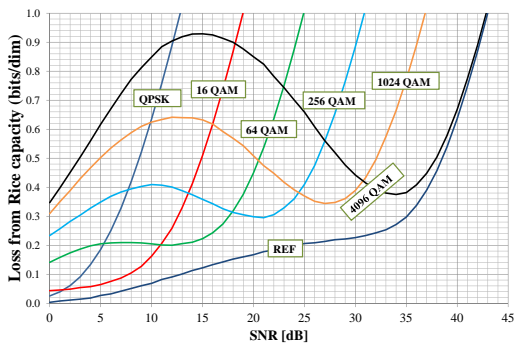}
          \vspace*{-0.5in}
        \caption{2SD-SH with two soft bits decoding per dimension }
        \label{fig:mouse}
    \end{subfigure}  \hfill
        \begin{subfigure}[b]{0.45\textwidth}
        \includegraphics[width=\textwidth]{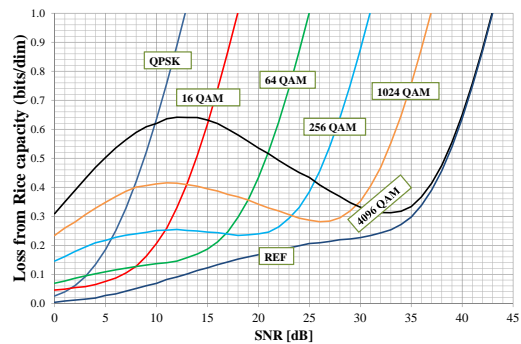}
          \vspace*{-0.5in}
        \caption{2SD-SH with three soft bits decoding per dimension }
        \label{fig:mouse}
    \end{subfigure}  \hfill
    \caption{Loss from the capacity over the Rician channel ($K=3$) for BICM and 2SD-SH schemes with various number of soft decoded bits per dimension.}\label{fig:RICE}
\end{figure}

\begin{figure}
    \centering
    \vspace*{-0.5in}
    \begin{subfigure}[b]{0.45\textwidth}
        \includegraphics[width=\textwidth]{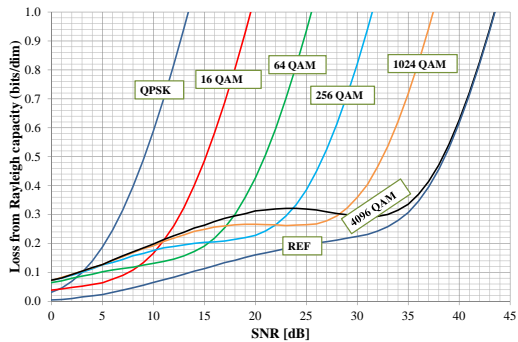}
        \vspace*{-0.5in}
        \caption{Loss from the capacity for BICM}
        \label{fig:gull}
    \end{subfigure}
    \hfill
    \begin{subfigure}[b]{0.45\textwidth}
        \includegraphics[width=\textwidth]{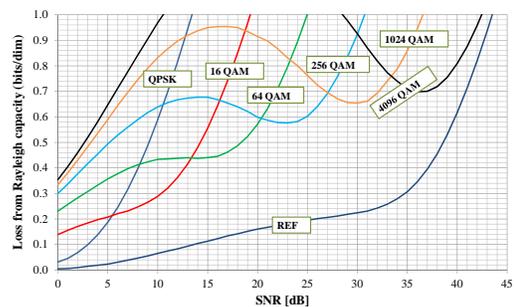}\\
          \vspace*{-0.5in}
        \caption{2SD-SH with one soft bit decoding per dimension }
        \label{fig:tiger}
    \end{subfigure}  \hfill
    \begin{subfigure}[b]{0.45\textwidth}
        \includegraphics[width=\textwidth]{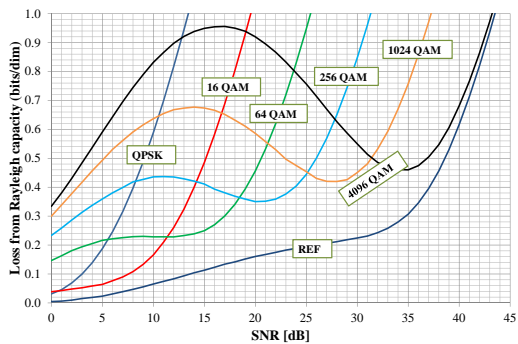}
          \vspace*{-0.5in}
        \caption{2SD-SH with two soft bits decoding per dimension }
        \label{fig:mouse}
    \end{subfigure}  \hfill
        \begin{subfigure}[b]{0.45\textwidth}
        \includegraphics[width=\textwidth]{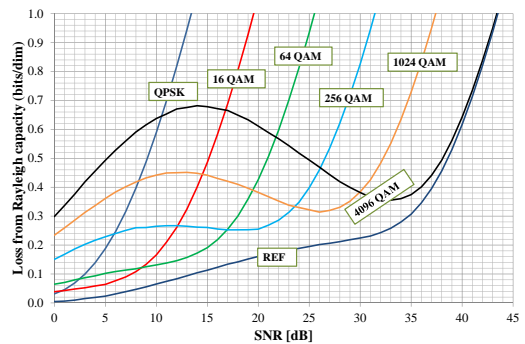}
          \vspace*{-0.5in}
        \caption{2SD-SH with three soft bits decoding per dimension }
        \label{fig:mouse}
    \end{subfigure}  \hfill
    \caption{Loss from the capacity over the Rayleigh channel for BICM and 2SD-SH schemes with various number of soft decoded bits per dimension.}\label{fig:RAY}
\end{figure}

The capacity of the Rician channel, with perfect CSI at the the receiver, can be calculated as
\cite{TSE}:
          \begin{multline}
          C=E_h\left\{\log\left(1+h^2 \frac{E_s}{N_0}\right)\right\}=
          \int_0^\infty 2(K+1) h e^{-K-(K+1)h^2}I_0(\alpha)\log\left(1+h^2 \frac{E_s}{N_0}\right)dh,
          \end{multline}
in which $\alpha = 2h\sqrt{K(K+1)} $ and $I_0(.)$ is the Bessel function. In Fig. \ref{fig:RICE} we
present the MI results for the Rician channel with $K=3$ [dB]. As before, the results for BICM, one
bit soft decoder, two bits soft stage 2SD-SH and three bits soft decoder 2SD-SH are respectively
plotted in Fig. \ref{fig:RICE}(a)-\ref{fig:RICE}(d). Notice that contrary to the AWGN, three soft
bits per dimension are needed in order to keep the capacity loss and code flexibility similar to
the BICM approach. This implies that only for 256-QAM and larger constellations the 2SD-SH scheme
is beneficial, particularly from the complexity point of view. These results do not vary much as a
function of $K$. Indeed, in Fig. \ref{fig:RAY} we plot the MI results also for the Rayleigh channel
(choosing $K=-100$ dB). The final conclusions do not change, although the loss with respect to the
BICM is slightly larger in comparison to the Rician channel.

\section{Simplification of LLR computations}\label{sec:LLRsimp}
At the receiver, first, we need to compute the log-likelihood ratios before feeding them to the
following soft iterative decoder. As it can be seen from equation (\ref{eq:comp}), the
computational complexity of LLR grows exponentially with the spectral efficiency and thus it
becomes dominant for large values of $m$. In this section we propose a method to simplify this
block.

Notice that symbol LLRs for the first stage, $\lambda(b_1)$, are computed by marginalizing the
channel conditional pdf w.r.t. the unknown sub-symbols relative to the second decoded stage $b_2$.
We consider a PAM modulation scheme with constellation points at positions
\[x = d\left[ B - \left(\frac{M - 1}{2} \right) \right]\quad B \in \left[ {0,M - 1} \right].\] We also assume the natural mapping assigning the least significant bits to the soft decoded stage as in \cite{Wachsmann_MLC_Hard}:
\[B= {B_2}{M_1} + {B_1}.\]

Under these assumptions, for the AWGN channel, the likelihoods can be computed as follows:
          \begin{eqnarray}
          \exp\left( \lambda \left( b_1 \right) \right) &\propto& \sum\limits_{b_2} p\left( Y|X(b_2,b_1)\right)\nonumber  \\
          &\propto & \sum_{b_2 = 0}^{M_2 - 1} e^{\left( - \frac{K_w}{2}\left\| Y - b_1d - b_2M_1d \right\|^2 \right)},
          \label{eq:llr}
          \end{eqnarray}
\noindent where $K_w=1/N_0$. As observed in \cite{Forney}, the likelihoods (\ref{eq:llr}) are
similar to wrapped and sampled Gaussian messages, where the number of Gaussian replicas is limited
(in the range $[0,M_2-1]$), instead of being unbounded.

When a \emph{binary} turbo decoder scheme is adopted at the soft decoded stage, symbol LLRs should
be successively converted to bit LLRs, with an additional complexity which is of order $O(m_1
2^{m_1})$.

The computational complexity of LLRs (\ref{eq:llr}) grows linearly with respect to the number of
constellation points, $M$, and independently from $M_1$. This complexity derives mostly from the
necessity of performing the sum of $M_2$ terms. By substituting the finite sum in (\ref{eq:llr})
with an unbounded sum, the resulting likelihoods become a truly wrapped and sampled Gaussian
distribution. In doing this, using the results in circular statistic, the wrapped Gaussian
distributions can be approximated by Von Mises distribution:
\[ \sum_{b_2 = -\infty}^{\infty} e^{\left( - \frac{K_w}{2}\left\| Y - b_1d - b_2M_1d \right\|^2 \right)}
\approx e^{K_v\cos\left(\frac{2\pi}{M_1d}( Y - b_1d ) \right)}
\]
with a suitable mapping between $K_v$ and $K_w$ (see for example \cite{ADBP_lett}). So that we can
use the following approximation for the log-likelihoods
\[
\lambda(b_1)\approx K_v \cos\left( \frac{2\pi}{M_1d}( Y - b_1d )\right),\label{eq:llrsim}
\]
with  a complexity that now depends only on $M_1$, which is typically very small given the results
in section \ref{sec:Results1}. Furthermore, since $Y$ appears at the argument of the cosine it can
be substituted with its ``wrapped'' version $Y'=Y\bmod (M_1d)$, thus requiring a smaller number of
bits for its representation.

\section{Embedding ADBP as the soft decoder in 2SD-SH}\label{sec:ADBP}
Non binary LDPC codes over rings of size $M$ have been introduced in \cite{Bennatan}. The main
advantage of such codes is close to capacity performance for rather short lengths as $M$ increases.
However, the believe propagation decoding algorithm becomes too complex by increasing $M$. ADBP has
been introduced in \cite{ADBP_lett} as an efficient iterative BP algorithm for decoding non binary
LDPC codes over rings. The peculiarity of this algorithm is that the computational complexity of
the message updating, at both variable and check nodes, \emph{is independent from} $M$. As a
consequence, it is possible to construct non binary decoders with complexity and memory
requirements independent from the cardinality of the alphabet $M$ \cite{awais2014vlsi}.

The messages that are processed by ADBP belong to the class of \emph{wrapped} and \emph{sampled}
Gaussian distributions (named named D-messages in \cite{ADBP_lett}). The main idea behind ADBP is
to represent each message only by two parameters, namely mean and variance ($Y$ and $K_w$ in
(\ref{eq:llr})). At each iteration  the algorithm needs to update these two parameters. Notice that
in on the other hand in  regular BP the messages are vectors of size $M$. The main result in
\cite{ADBP_lett} shows that by constructing LDPC codes over rings, that prevent the use of
multiplications by coefficients different from $\pm 1$ in the code construction, one can force the
BP messages to stay approximately within the class of D-messages. The ADPB then defines the correct
updating of the parameters ($Y$ and $K_w$) to be performed at the variable and check nodes.

In order to use ADPB, also the input \emph{channel} messages should be D-messages. It is
interesting to notice that messages in equation (\ref{eq:llr}) actually belong to the D-message
class and no further approximations are necessary. Therefore, the ADBP algorithm seems to be the
natural candidate for performing soft decoding of the first stage of the considered 2SD-SH. Notice
that ADBP does not require the LLR computation, neither the conversion of symbol to bit LLR, and
therefore the detection complexity disappears.

As we have mentioned, the ADBP requires that the weights of the edges in the factor graph of the
LDPC to be chosen from the set $\{ \pm 1 \}$. Thus, one cannot use the same code designs which has
been proposed in the literature previously. A natural question is whether it is possible to design
such codes with performances competitive to those that can be obtained without having the stringent
condition on weights. This problem is open in general and is the main subject of our future
research.

\section{Simulation Results}\label{sec:simul}
\begin{figure}[tb]
	\begin{center}
	\vspace*{-0.4in}
	\includegraphics[angle=0,width=\the\hsize]{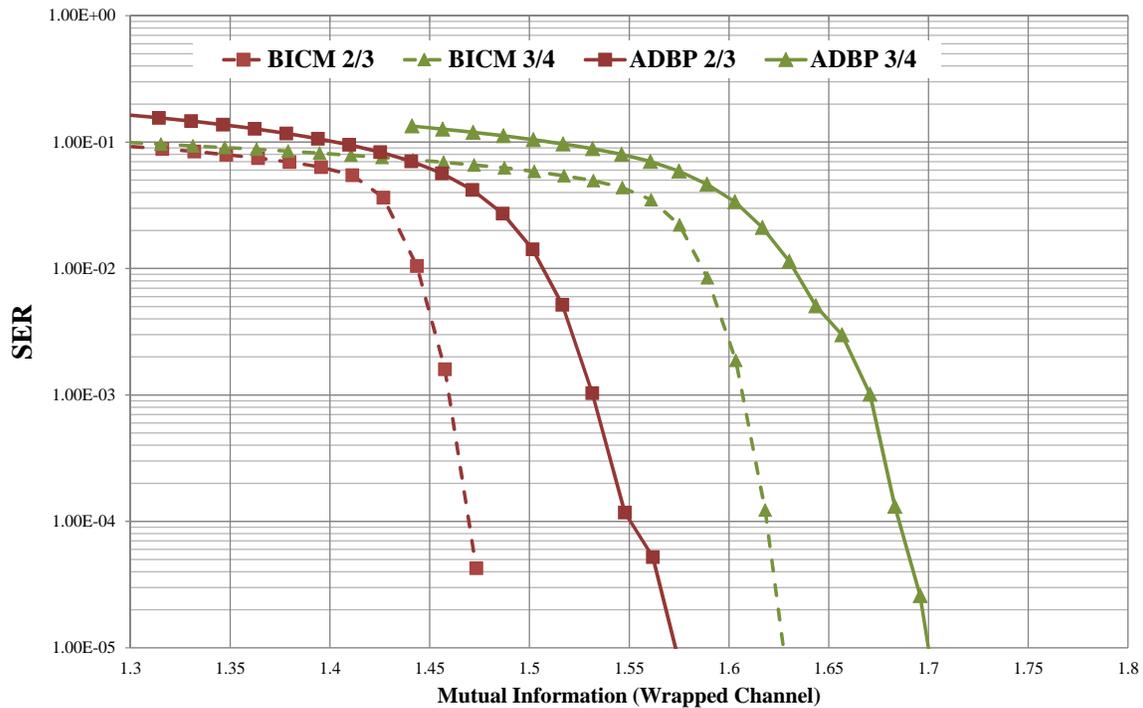}
	\vspace*{-0.4in}
	\caption {\label{Fig:23adbpvsBICM}  Comparison between
	the Symbol Error Rate performances relative to two coding solutions for the soft stage of the 2SD-SH,
versus the MI of the correspondent wrapped channel.}
	\end{center}
\end{figure}

In Figure~\ref{Fig:23adbpvsBICM} we show a comparison between the performances relative to two
coding solutions for the soft stage of the 2SD-SH. We have chosen two different rates, namely $R_1
= 2/3$ and $R_2 = 3/4$.
In the first ``classical'' coding scheme a rate 2/3 SCCC binary encoder is used to generate the
bits that are mapped to a 4-PAM constellation with Gray mapping and then sent through a
\emph{wrapped} Gaussian channel. In the second scheme a quaternary regular LDPC encoder with rate
2/3 (variable degree $d_v=3$) is used to directly generate the indexes of the 4-PAM constellations.
The performances are reported versus the MI (in bits) of the considered wrapped channel. An ideal
capacity achieving coding scheme with the same code parameter would have a SER dropping to zero for
a mutual information of the wrapped channel equal to $2/3\times 2=1.333$. The gap of the reported
performance curves with respect to this value thus represents the throughput loss due to the
suboptimal code choice. The ADBP based scheme have a loss of about 0.1 bit w.r.t. the more
conventional approach. In the same figure we also report the results for a coding scheme with rate
3/4 (green curves with triangular markers). In this case the ideal capacity achieving coding scheme
would have a SER dropping to zero for mutual information value of $3/4 \times 2 = 1.5$. Notice that
since the total throughput of the two stage system is given by the sum of the throughput of the
hard and soft decoding stages, and hard decoding throughput is dominant for large $m$, the relative
throughput loss due to suboptimal soft decoding becomes negligible when $m$ increases.

\begin{figure}[tb]
	\begin{center}
	\vspace*{-0.4in}
	\includegraphics[angle=0,width=\the\hsize]{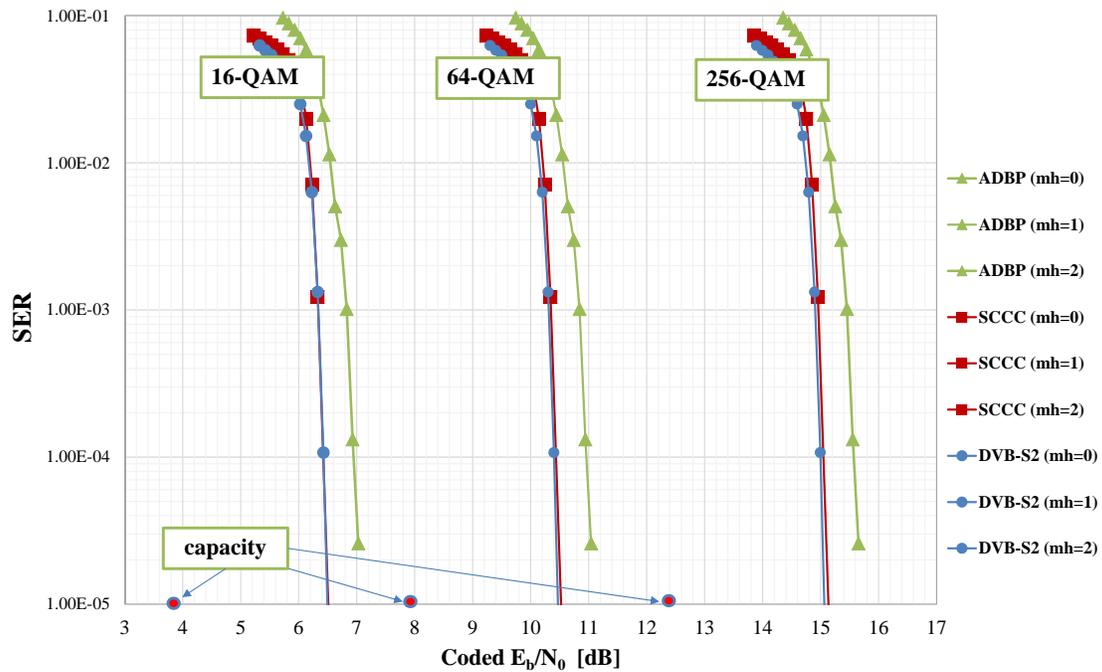}
	\vspace*{-0.5in}
	\caption {\label{Fig:SER_ADPB} Symbol error rate of two layer coding scheme with simplified detection and decoding using the ADBP algorithm.}
	\end{center}
\end{figure}

In Fig. \ref{Fig:SER_ADPB} we show the simulation results for a full 2SD-SH scheme embedding in the
first stage a non binary LDPC (with associated ADBP decoder) and more classical rate 3/4 binary
turbo-like encoders. We show the SER for three  constellation sizes (16, 64, and 256QAM),
corresponding to efficiencies of 1.5, 2.5, and 3.5 bits per dimension. In all cases two bits per
dimension are soft decoded and the number of hard decoded bits per dimension, denoted by $mh$, is
reported in the legend for each curve. The scheme with ADBP adopts a regular rate 3/4 quaternary
LDPC code ($d_v=3, d_c=12$).  The reference turbo-like encoders  are  the DVB-S2 LDPC code or a
Turbo SCCC code. The 2SD-SH based on ADBP shows around 0.5 loss with respect to the state of the
art LDPC or SCCC codes. This loss is due to the weakness of the regular LDPC code used rather than
to the suboptimality of the ADBP decoder, which actually provides performance very close to the
optimal but more complex non binary BP decoder (\cite{awais2014vlsi}).

\section{Conclusions}\label{sec:Conclusion}
In this paper we analyzed a two level coding scheme with second  stage begin hard decoded,
named 2SD-SH. We show that it is possible to construct a flexible two stage coding scheme achieving
performances close to the theoretical limits for all desired spectral efficiencies by pairing a
powerful coding scheme with a simpler algebraic code encoding the majority of modulation bits.
These conclusions confirm and  extend what has been observed in literature
\cite{Wachsmann_MLC_Hard}, \cite{Forney}. Moreover, we have shown that the number of required soft
decoded bits to preserve the flexibility of the two stage encoder is two per dimension over the
AWGN channel. Over the Rayleigh channel, the needed number of bits per dimension to be soft decoded
is three.

We then proposed  a simplification of the computation of the LLR for the soft decoded stage
allowing to further decrease the complexity. The resulting two stage scheme is better in term of
flexibility and receiver complexity with respect to BICM that is currently adopted in many
standards as high spectral efficient coding scheme.

Finally we have shown that the recently introduced ADBP algorithm is a natural candidate to perform
the soft decoding stage for our proposed scheme. The adoption of this approach allows to  remove
the need of LLR computation, reducing even further the complexity. Simulation results show that a
solution based on ADBP has performance competitive but worse than those that can be obtained
with a more conventional soft decoding approach.
\balance

\section*{Acknowledgment}
The research described in this paper was partly carried out under contract with Ericsson AB. This
work is partially supported by the NEWCOM\# network of excellence, funded by the European
Community. 

Farbod Kayhan is supported by the Luxembourg National Research Fund under CORE Junior project: C16/IS/11332341 (ESSTIMS).

%

\begin{thebibliography}{10}
\providecommand{\url}[1]{#1}
\csname url@samestyle\endcsname
\providecommand{\newblock}{\relax}
\providecommand{\bibinfo}[2]{#2}
\providecommand{\BIBentrySTDinterwordspacing}{\spaceskip=0pt\relax}
\providecommand{\BIBentryALTinterwordstretchfactor}{4}
\providecommand{\BIBentryALTinterwordspacing}{\spaceskip=\fontdimen2\font plus
\BIBentryALTinterwordstretchfactor\fontdimen3\font minus
  \fontdimen4\font\relax}
\providecommand{\BIBforeignlanguage}[2]{{%
\expandafter\ifx\csname l@#1\endcsname\relax
\typeout{** WARNING: IEEEtran.bst: No hyphenation pattern has been}%
\typeout{** loaded for the language `#1'. Using the pattern for}%
\typeout{** the default language instead.}%
\else
\language=\csname l@#1\endcsname
\fi
#2}}
\providecommand{\BIBdecl}{\relax}
\BIBdecl

\bibitem{BICM}
G.~Caire, G.~Taricco, and E.~Biglieri, ``Bit-interleaved coded modulation,''
  \emph{Information Theory, IEEE Transactions on}, vol.~44, no.~3, pp.
  927--946, 1998.

\bibitem{Imai_Hirakawa}
H.~Imai and S.~Hirakawa, ``A new multilevel coding method using
  error-correcting codes,'' \emph{Information Theory, IEEE Transactions on},
  vol.~23, no.~3, pp. 371--377, 1977.

\bibitem{Wachsmann_MLC_Hard}
U.~Wachsmann, R.~Fischer, and J.~Huber, ``Multilevel coding: use of hard
  decisions in multistage decoding,'' in \emph{Proceedings of the annual
  allerton conference on communication control and computing}, vol.~35.\hskip
  1em plus 0.5em minus 0.4em\relax Citeseer, 1997, pp. 966--975.

\bibitem{Wachsmann_MLC}
U.~Wachsmann, R.~F. Fischer, and J.~B. Huber, ``Multilevel codes: Theoretical
  concepts and practical design rules,'' \emph{Information Theory, IEEE
  Transactions on}, vol.~45, no.~5, pp. 1361--1391, 1999.

\bibitem{Philippa}
P.~A. Martin and D.~P. Taylor, ``On multilevel codes and iterative multistage
  decoding,'' \emph{Communications, IEEE Transactions on}, vol.~49, no.~11, pp.
  1916--1925, 2001.

\bibitem{IHID}
M.~A. Naim, J.~P. Fonseka, and E.~M. Dowling, ``Improved hard iterative
  decoding {(IHID)} of multilevel codes,'' \emph{Communications Letters, IEEE},
  vol.~17, no.~7, pp. 1443--1446, 2013.

\bibitem{DRM}
V.~Fischer, A.~Kurpiers, and F.~Kulla, ``Improved multistage decoding of
  multilevel codes for digital radio mondiale (drm),'' in \emph{8th IEEE
  International Symposium on Consumer Electronics 2004}, 2004.

\bibitem{Forney}
G.~D. Forney~Jr, M.~D. Trott, and S.-Y. Chung, ``Sphere-bound-achieving coset
  codes and multilevel coset codes,'' \emph{Information Theory, IEEE
  Transactions on}, vol.~46, no.~3, pp. 820--850, 2000.

\bibitem{ADBP_lett}
G.~Montorsi, ``Analog digital belief propagation,'' \emph{Communications
  Letters, IEEE}, vol.~16, no.~7, pp. 1106--1109, 2012.

\bibitem{blacksea}
G.~Montorsi and F.~Kayhan, ``Analog digital belief propagation and its
  application to multi stage decoding systems,'' in \emph{2015 IEEE
  International Black Sea Conference on Communications and Networking
  (BlackSeaCom),}, May 2015, pp. 82--86.

\bibitem{TSE}
D.~Tse and P.~Viswanath, \emph{Fundamentals of Wireless Communication}.\hskip
  1em plus 0.5em minus 0.4em\relax Cambridge University Press, 2005.

\bibitem{Bennatan}
A.~Bennatan and D.~Burshtein, ``Design and analysis of nonbinary {LDPC} codes
  for arbitrary discrete-memoryless channels,'' \emph{Information Theory, IEEE
  Transactions on}, vol.~52, no.~2, pp. 549--583, 2006.

\bibitem{awais2014vlsi}
M.~Awais, G.~Masera, M.~Martina, and G.~Montorsi, ``{VLSI} implementation of a
  non-binary decoder based on the analog digital belief propagation,''
  \emph{Signal Processing, IEEE Transactions on}, vol.~62, no.~15, pp.
  3965--3975, Aug 2014.

\end{thebibliography}
\end{document}